\documentclass[journal,transmag]{IEEEtran}
\usepackage{cite}
\usepackage{url}
\usepackage[normalem]{ulem}

\ifCLASSINFOpdf
  \usepackage[pdftex]{graphicx}
\fi
\usepackage{xcolor}

\usepackage[cmex10]{amsmath}
\usepackage{amssymb}
\def\apj{Astrophys. J.}
\def\apjl{Astrophys. J. Lett.}
\def\apjs{Astrophys. J. Supp. Ser. }

\def\prd{Phys. Rev. D.}

\begin{document}
%
\title{Massive Computation for Understanding \\ Core-Collapse Supernova Explosions}


\author{\IEEEauthorblockN{Christian D. Ott\IEEEauthorrefmark{1,2}}
  \IEEEauthorblockA{\IEEEauthorrefmark{1}TAPIR, Walter Burke Institute for Theoretical Physics, California Institute of Technology, Pasadena, CA, 91125, USA} 
\IEEEauthorrefmark{2}Yukawa Institute for Theoretical Physics, Kyoto University, Kyoto, Japan
}

\markboth{C. D. Ott, Manuscript for Computing in Science and Engineering (CiSE), dated: \today}%
{Shell \MakeLowercase{\textit{et al.}}: Bare Demo of IEEEtran.cls for Journals}
%



\IEEEtitleabstractindextext{%
  \begin{abstract}
    How do massive stars explode? Progress toward the answer is driven
    by increases in compute power.  Petascale supercomputers are
    enabling detailed three-dimensional simulations of core-collapse
    supernovae. These are elucidating the role of fluid instabilities,
    turbulence, and magnetic field amplification in supernova engines.
  \end{abstract}

}

\maketitle

\IEEEdisplaynontitleabstractindextext

%
\IEEEpeerreviewmaketitle

\section{Introduction}

Core-collapse supernova explosions come from stars more massive than
$\sim$$8-10$ times the mass of the Sun. Ten core-collapse supernovae
explode per second in the universe, automated astronomical surveys
discover multiple per night, and one or two explode per century in the
Milky Way. Core-collapse supernovae outshine entire galaxies in
photons for weeks and output more power in neutrinos than the combined
light output of all other stars in the universe, for tens of
seconds. These explosions pollute the interstellar medium with the
ashes of thermonuclear fusion. From these elements, planets form and
life is made. Supernova shock waves stir the interstellar gas, trigger
or shut off the formation of new stars, and eject hot gas from
galaxies. At their centers, a strongly gravitating compact remnant, a
neutron star or a black hole, is formed.

\begin{figure}[h!]
  \centering
\includegraphics[width=0.97\columnwidth]{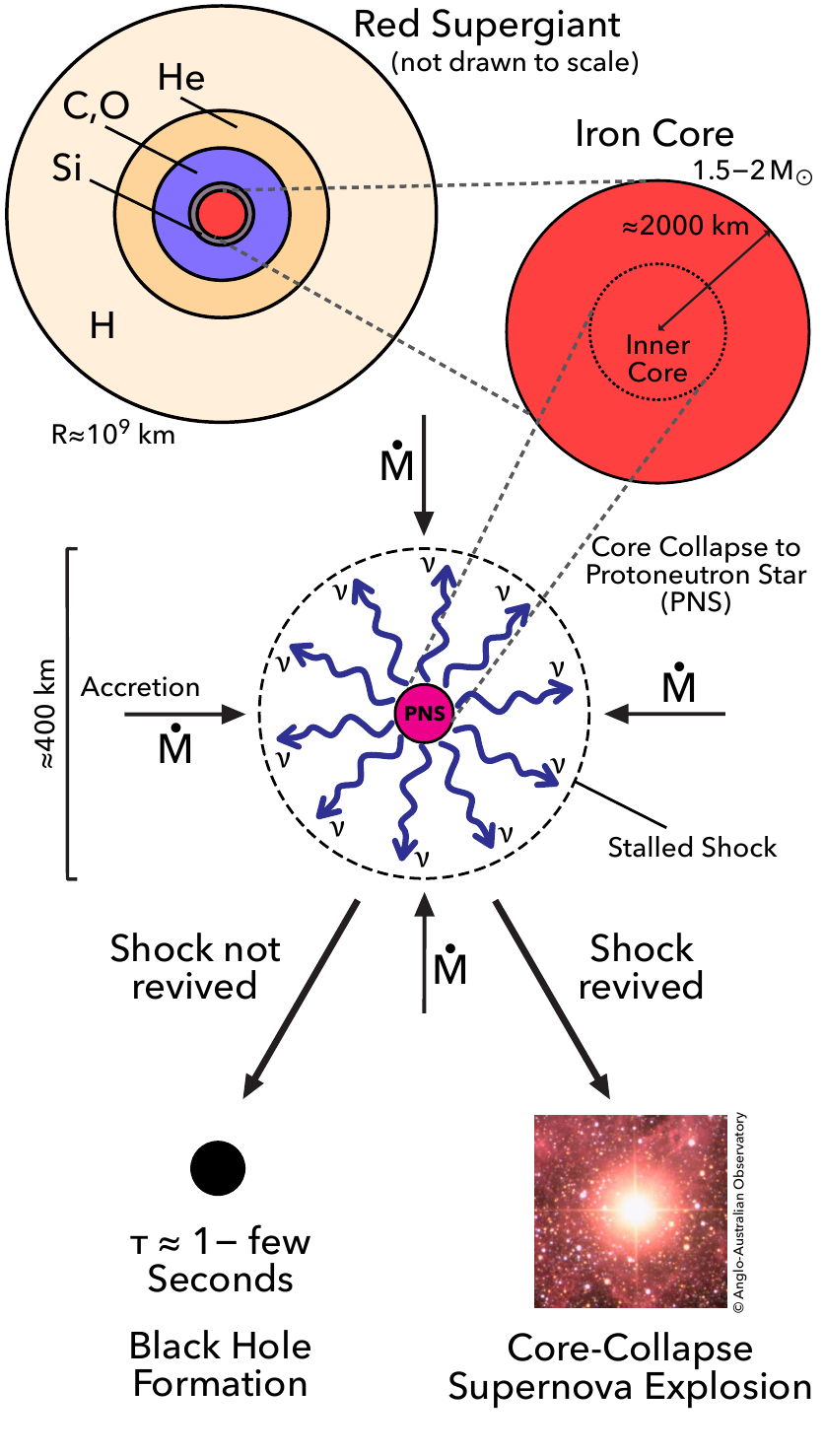}
\caption{Schematic of core collapse and its simplest
  outcomes. The image shows SN 1987A, which exploded in the Large
  Magellanic Cloud.}
\label{fig:collapse}
\end{figure}

As the name alludes, the explosion is preceded by collapse of a
stellar core. At the end of its life, a massive star has a core
composed mostly of iron-group nuclei. The core is surrounded by an
onion-skin structure of shells dominated by successively lighter
elements. Nuclear fusion is still ongoing in the shells, but the iron
core is inert. The electrons in the core are relativistic and
degenerate. They provide the lion's share of the pressure support
stabilizing the core against gravitational collapse. In this, the iron
core is very similar to a white dwarf star, the end product of
low-mass stellar evolution.  Once the iron core exceeds its maximum
mass (the so-called effective Chandrasekhar mass of $\sim$$1.5-2$
  solar masses [$M_\odot$]), gravitational instability sets in.
  Within a few tenths of a second, the inner core collapses from a
  central density of $\sim$$10^{10}\,\mathrm{g\,cm}^{-3}$ to a density
  comparable to that in an atomic nucleus ($\gtrsim 2.7\times
  10^{14}\,\mathrm{g\,cm}^{-3}$). There, the repulsive part of the
  nuclear force causes a stiffening of the equation of state (EOS; the
  pressure--density relationship). The inner core first overshoots
  nuclear density, then rebounds (``bounces'') into the still
  collapsing outer core. The inner core then stabilizes and forms the
  inner regions of the newborn protoneutron star. The {hydrodynamic}
  supernova shock is created at the interface of inner and outer
  core. First, the shock moves outward dynamically. It then quickly
  loses energy by work done breaking up infalling iron-group nuclei
  into neutrons, protons, and alpha particles. The copious emission of
  neutrinos from the hot ($T \sim 10 \,\mathrm{MeV} \simeq
  10^{11}\,\mathrm{K}$) gas further reduces energy and pressure behind
  the shock. The shock stalls and turns into an accretion shock: the
  ram pressure of accretion of the star's outer core balances the
  pressure behind the shock.

The \emph{supernova mechanism} must revive the stalled shock to drive
a successful core-collapse supernova explosion. Depending on the
structure of the progenitor star, this must occur within one to a few
seconds of core bounce. Otherwise, continuing accretion pushes the
protoneutron star over its maximum mass ($\sim$$2-3\,M_\odot$), which
results in the formation of a black hole and no supernova explosion.

If the shock is successfully revived, it must travel through the outer
core and the stellar envelope before it breaks out of the star and
creates the spectacular explosive display observed by astronomers on
Earth. This may take more than a day for a red supergiant star (e.g.,
like Betelgeuse, a $\sim$$20\,M_\odot$ star in the constellation
Orion) or just tens of seconds for a star that has been stripped of
its extended hydrogen-rich envelope by a strong stellar wind or mass
exchange with a companion star in a binary system.

The photons observed by astronomers are emitted extremely far from the
central regions. They carry information on the overall energetics, the
explosion geometry, and on the products of explosive nuclear burning
that is triggered by the passing shock wave. They can, however, only
provide weak constraints on the inner workings of the
supernova. Direct observational information on the supernova mechanism
can be gained only from neutrinos and gravitational waves that are
emitted directly in the supernova core. Detailed computational models
are required for gaining theoretical insight and for making
predictions that can be contrasted with future neutrino and
gravitational-wave observations from the next core-collapse supernova
in the Milky Way.

\section{Supernova Energetics and Mechanisms}

Core-collapse supernovae are ``gravity bombs.'' The energy reservoir
from which any explosion mechanism must draw is the gravitational
energy released in the collapse of the iron core to a neutron star:
$\sim$$3\times10^{53}\,\mathrm{erg}$ ($3\times 10^{46}\,\mathrm{J}$),
a mass-energy equivalent of $\sim$$0.15M_\odot c^2$. A fraction of
this tremendous energy is stored initially as heat (and rotational
kinetic energy) in the protoneutron star and the rest comes from its
subsequent contraction. Astronomical observations, on the other hand,
show the typical core-collapse supernova explosion energy to be in the
range $10^{50}-10^{51}\,\mathrm{erg}$. \emph{Hypernova} explosions may
have up to $10^{52}\,\mathrm{erg}$, but they make up $\lesssim$1\% of
all core-collapse supernovae. A small subset of hypernovae are
associated with gamma-ray bursts.

Where is all the gravitational energy going that does not contribute
to the explosion energy? The answer is: Neutrinos. Antineutrinos and
neutrinos of all flavors carry away $\gtrsim 99\%$ ($\gtrsim 90\%$ in
the hypernova case) of the available energy over
$\mathcal{O}(10)\,\mathrm{s}$ as the protoneutron star cools and
contracts. This was first theorized and then later observationally
confirmed with the detection of neutrinos from SN~1987A, the most
recent core-collapse supernova in the Milky Way vicinity.

\begin{figure}[t]
  \centering
\includegraphics[width=\columnwidth]{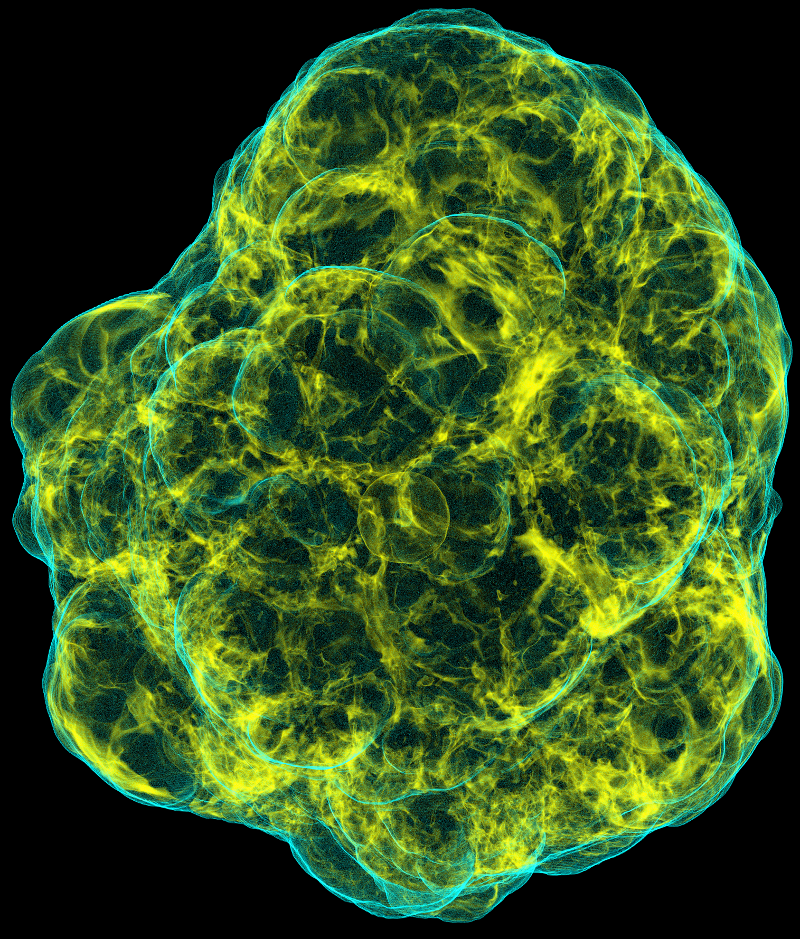}
\caption{Volume rendering of the specific entropy in the core of a
  neutrino-driven core-collapse supernova at the onset of explosion.
  Based on the 3D general-relativistic simulations of \cite{ott:13a}
  and rendered by Steve Drasco (Cal Poly San Luis Obispo). Specific
  entropy is a preferred quantity for visualization, since in the core
  of a supernova, it typically ranges from $\sim$1 to $\sim$20 units
  of Boltzmann's constant $k_\mathrm{B}$ per baryon.  Shown is
  the large-scale asymmetric shock front and a layer of hot expanding
  plumes behind it. The physical scale is roughly $600 \times
  400\,\mathrm{km}$.}
\label{fig:full3d}
\end{figure}

Since neutrinos dominate the energy transport through the supernova,
they might quite naturally have something to do with the explosion
mechanism. The \emph{neutrino mechanism}, in its current form, was
proposed by Bethe \& Wilson \cite{bethewilson:85}. In this mechanism,
a fraction ($\sim$$5\%$) of the outgoing electron neutrinos and
antineutrinos is absorbed in a layer between protoneutron star and the
stalled shock. In the simplest picture, this neutrino \emph{heating}
increases the thermal pressure behind the stalled shock. Consequently, the
dynamical pressure balance at the accretion shock is violated and a
runaway explosion is launched.

The neutrino mechanism fails in spherical symmetry (1D, e.g.,
\cite{janka:12a}), but is very promising in multiple dimensions
(axisymmetry [2D], 3D).  This is due largely to multi-D hydrodynamic
instabilities that break spherical symmetry (see
Figure~\ref{fig:full3d} for an example), increase the neutrino
mechanism's efficiency, and facilitate explosion.  I will discuss this
in more detail later in this article. The neutrino mechanism is
presently favored as the mechanism driving most core-collapse
supernova explosions (see \cite{janka:12a} for
a recent review).

\begin{figure}[t]
  \centering
\includegraphics[width=0.9\columnwidth]{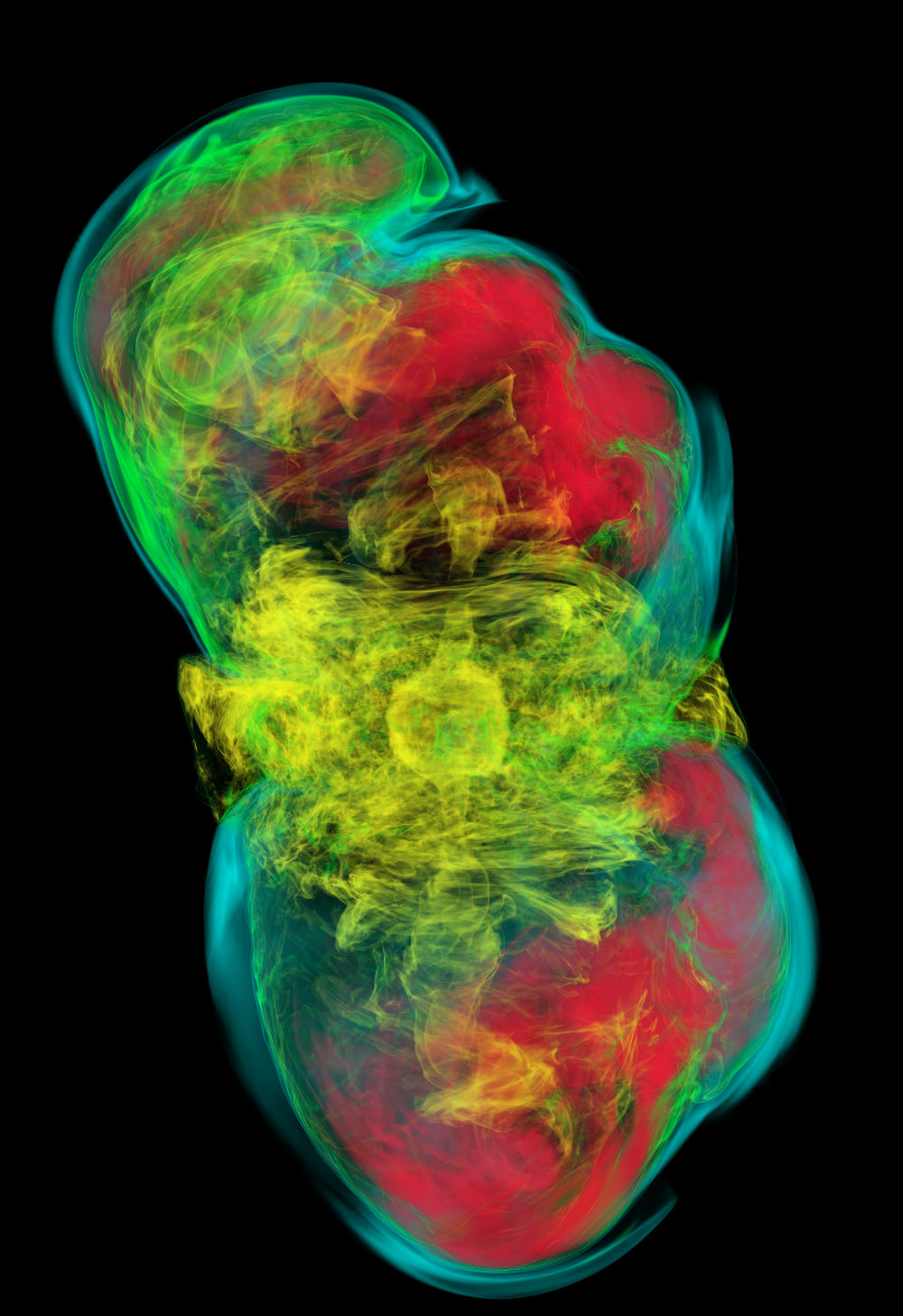}
\caption{Volume rendering of the specific entropy in
  the core of a magnetorotational core-collapse supernova. Bluish
  colors indicate low entropy, red colors high entropy, and green and
  yellow intermediate entropy. The vertical is the axis of rotation
  and shown is a region of $\sim$$1600 \times 800\,\mathrm{km}$.
  The ultra-strong toroidal magnetic field surrounding the the protoneutron
  star pushes hot plasma out along the rotation axis. The distorted, 
  double-lobe structure is due to an MHD kink instability akin those seen
  in Tokamak fusion experiments.   Used with permission
  from M\"osta~\emph{et al.}~2014~\cite{moesta:14b}.}
\label{fig:mhd2014}
\end{figure}

Despite its overall promise, the neutrino mechanism is very
inefficient. Only $\lesssim 5\%$ of the outgoing total electron
neutrino and antineutrino luminosity is deposited behind the stalled
shock at any moment and much of this deposition is lost again as
heated gas flows down, leaves the heating region, and settles onto the
protoneutron. The neutrino mechanism may (barely) be able to power
ordinary core-collapse supernovae, but it cannot deliver hypernova
explosion energies or account for gamma-ray bursts.

An alternative mechanism that may be part of the explanation for such
extreme events is the \emph{magnetorotational mechanism}, first
suggested by Bisnovatyi-Kogan \cite{bisno:70} and LeBlanc \& Wilson
\cite{leblanc:70}. In its modern form, a very rapidly spinning core
collapses to a protoneutron star with a spin period of only
$\sim$$1\,\mathrm{millisecond}$. Its core is expected to be spinning
uniformly, but its outer regions will be extremely differentially
rotating. These are ideal conditions for the \emph{magnetorotational
  instability} (MRI,\cite{balbus:91}) to operate, amplify any seed
magnetic field, and drive magnetohydrodynamic (MHD) turbulence. If a
dynamo process is present, an ultra-strong large-scale (globally
  ordered) magnetic field is built up. This makes the protoneutron
  star a \emph{protomagnetar}. Provided this occurs, magnetic
pressure gradients and hoop stresses could lead to outflows along the
axis of rotation.  The MRI's fastest growing mode has a small
wavelength and is extremely difficult to resolve numerically.

Because of this, all simulations of the magnetorotational mechanism to
date have simply made the assumption that a combination of MRI and
dynamo is operating. They then ad-hoc imposed a strong large-scale
field as an initial condition. In 2D simulations, collimated jets
  develop along the axis of rotation. In 3D, the jets are unstable and
  a more complicated explosion geometry develops \cite{moesta:14b}, as
  shown in Figure~\ref{fig:mhd2014}. Nevertheless, even in 3D, an
  energetic explosion could potentially be powered.

The magnetorotational mechanism requires one special property of the
progenitor star: rapid core rotation. Presently, stellar evolution
theory suggests that the cores of most massive stars should be slowly
spinning. However, there may be exceptions of rapidly spinning cores at
just about the right occurrence rate to explain hypernovae and long
gamma-ray bursts.

Besides the neutrino mechanism and the magnetorotational mechanism, a
number of other explosion mechanisms have been proposed. I direct the
interested reader to the more extensive review by \cite{janka:12a}.

\section{A Multi-Scale, Multi-Physics, Multi-Dimensional Computational Challenge}

The core-collapse supernova problem is highly complex, inherently
non-linear, and involves many branches of (astro)physics. Only limited
progress can be made with analytic or perturbative
methods. Computational simulation is a powerful means for gaining
theoretical insight and for making predictions that could be tested
with astronomical observations of neutrinos, gravitational waves, and
electromagnetic radiation.

Core-collapse supernova simulations are time evolution simulations --
starting from initial conditions, the matter, radiation, and
gravitational fields are evolved in time. In the case of time-explicit
evolution, the numerical timestep is limited by causality, controlled
by the speed of sound in Newtonian simulations, and the speed of light
in general-relativistic simulations. Because of this, an increase in
the spatial resolution by a factor of two corresponds to a decrease in
the time step by a factor of two. Hence, in a 3D simulation, the
computational cost scales with the fourth power of resolution.

\subsection{Multi Scale}

Taking the red supergiant in Figure~\ref{fig:collapse} as an example,
a complete core-collapse supernova simulation that follows the shock
to the stellar surface, would have to cover dynamics on a physical
scale from $\sim$$10^9\,\mathrm{km}$ (stellar radius) down to
$\sim$$0.1\,\mathrm{km}$ (the typical scale over which structure and
thermodynamics of the protoneutron star change).  These ten orders of
magnitude in spatial scale are daunting. In practice, reviving the
shock and tracking its propagation to the surface can be treated as
(almost) independent problems. If our interest is on the shock revival
mechanism, we need to include the inner $\sim$$10,000\,\mathrm{km}$ of
the star. Since information about core collapse is communicated to
overlying layers with the speed of sound, stellar material at greater
radii will not ``know'' that core collapse has occurred before it is
hit by the revived expanding shock.

Even with only five decades in spatial scale, some form of grid
refinement or adaptivity is called for: a 3D finite-difference grid
with an extent of $10,000\,\mathrm{km}$ symmetric about the origin
with uniform $0.1\,\mathrm{km}$ cell size would require 57 PB of RAM
to store a single double precision variable. Many tens to hundreds of
3D variables are required. Such high uniform resolution is not only
currently impossible but also unnecessary. Most of the resolution is
needed near the protoneutron star and in the region behind the stalled
shock.  The near-free-fall collapse of the outer core can be simulated
with much lower resolution.

Because of the broad range of physics involved (see below) and the
limited available compute power, early core-collapse supernova
simulations were spherically symmetric (1D). 1D simulations often
employ a Lagrangian, comoving mass coordinate discretization. This
grid  can be set up to provide just the right resolution where and
when needed or can be dynamically re-zoned (an adaptive mesh
refinement [AMR] technique). Other 1D codes discretize in the Eulerian
frame and use a fixed grid whose cells are radially stretched using
geometric progression.

In 2D simulations, Eulerian, geometrically-spaced fixed spherical
grids are the norm, but some codes use cylindrical coordinates and
AMR. Spherical grids, already in 2D, suffer from a coordinate
singularity at the axis that can lead to numerical artifacts. In 3D,
they become even more difficult to handle and their focusing grid
lines impose a severe timestep constraint near the origin. Some 3D
codes still use a spherical grid, while many others employ Cartesian
AMR grids. Recent innovative approaches use so-called multi-block
grids with multiple curvilinear touching or overlapping logically
Cartesian ``cubed-sphere'' grids (e.g., \cite{wongwathanarat:10}).

\subsection{Multi Physics}
Core-collapse supernovae are very rich in physics. All fundamental
forces are involved and essential to the core collapse
phenomenon. These forces are probed under conditions that are
impossible (or exceedingly difficult) to create in earthbound
laboratories.

Gravity drives the collapse and provides the energy reservoir. It is
so strong near the protoneutron star that general relativity becomes
important and its Newtonian description does not suffice.  The
electromagnetic force describes the interaction of the dense,
hot magnetized, perfectly conducting plasma and the photons that
provide thermal pressure and make the supernova light. The weak force
governs the interactions of neutrinos and the strong (nuclear) force
is essential in the nuclear EOS and nuclear reactions. 

All this physics occurs at the microscopic, per particle level.
Fortunately, the continuum assumption holds, allowing us to describe
core-collapse supernovae on a macroscopic scale by a coupled set of
systems of non-linear partial differential equations (PDEs):

\begin{figure}[t]
  \centering
\includegraphics[width=0.8\columnwidth]{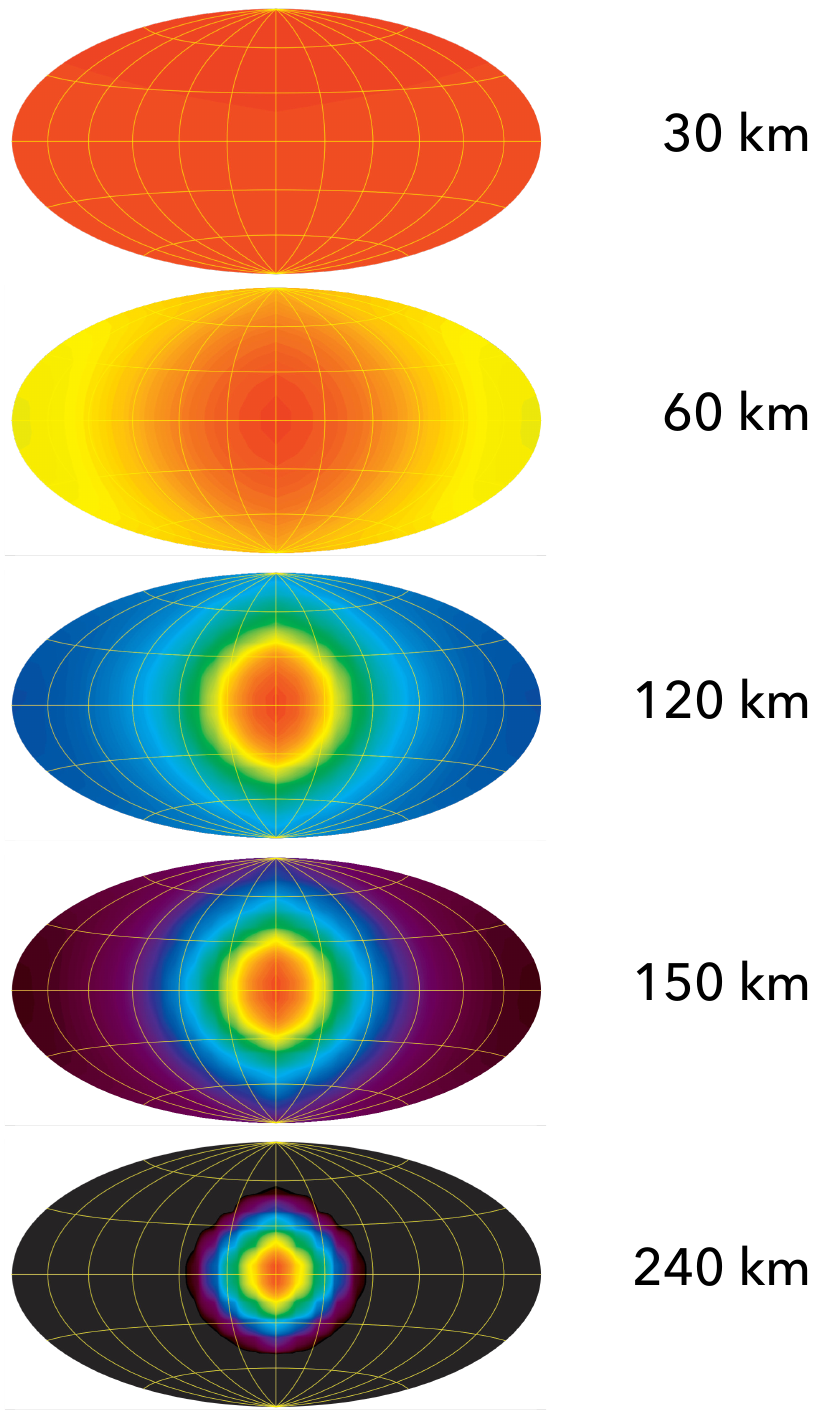}
\caption{Map projections of the momentum-space neutrino radiation
  field (for $\nu_e$ at an energy of 16.3\,MeV) going outward radially
  (from top to bottom) on the equator of a supernova core.  Generated
  using the simulation results of \cite{ott:08}. Inside the
  protoneutron star ($R \lesssim 30\,\mathrm{km}$) neutrinos and
  matter are in equilibrium and the radiation field is isotropic. It
  becomes more and more forward peaked as the neutrinos decouple and
  become free streaming. Handling the transition from slow diffusion to
  free streaming correctly requires angle-dependent radiation transport, which
  is a 6+1 D problem and computationally extremely challenging.}
\label{fig:momspace}
\end{figure}

\begin{itemize}
\item \textbf{(Magneto)hydrodynamics} (MHD). The stellar plasma is in
  local thermodynamic equilibrium, essentially perfectly conducting,
  and essentially inviscid (though neutrinos may provide some shear
  viscosity in the protoneutron star). The ideal, inviscid MHD
  approximation is appropriate under these conditions. The MHD
  equations are hyperbolic and can be written in flux-conservative
  form with source terms that do not include derivatives of the MHD
  variables. They are typically solved with standard time-explicit
  high-resolution shock capturing methods that exploit the
  characteristic structure of the equations (e.g.,
  \cite{toro:99}). Special attention must be paid to preserving the
  divergence-free property of the magnetic field. The MHD equations
  require an EOS as a closure (see below).

  \hspace{1em} Unless ultra-strong ($B\gtrsim 10^{15}\,\mathrm{G}$),
  magnetic fields have little effect on the supernova dynamics and
  thus are frequently neglected. Since strong gravity and velocities
  up to a few tenths of the speed of light are involved, the MHD
  equations are best solved in a general-relativistic
  formulation. General-relativistic MHD is computationally
  particularly expensive, because the conserved variables are not the
  primitive variables (density, internal energy / temperature,
  velocity, chemical composition). The latter are needed for the
    EOS and enter flux terms. After each update, they must be
    recovered from the conserved variables via multi-dimensional root
    finding.

  \vspace*{0.5em}
  
\item \textbf{Gravity}.  Deviations in the strength of the
  gravitational acceleration between Newtonian and
  general-relativistic gravity are small in the precollapse core, but
  become of order $10-20\%$ in the protoneutron star phase. In the
  case of black hole formation, Newtonian physics breaks down
  completely. General relativistic gravity is included at varying
  levels in simulations.  Some neglect it completely and solve the
  linear elliptic Newtonian Poisson equation to compute the
  gravitational potential. This is done using direct multigrid methods
  or integral multipole expansion methods. Some codes modify the
  monopole term in the latter approach to approximate general
  relativistic effects.

  \hspace{1em} Including full general relativity is more challenging,
  in particular in 2D and 3D, since there general relativity has
  radiative degrees of freedom (gravitational waves). An entire
  subfield of gravitational physics, \emph{numerical relativity},
  spent nearly five decades looking for ways to solve Einstein's
  equations on computers (see \cite{baumgarte:10book} for a
  comprehensive introduction).  In general relativity, changes in the
  gravitational field propagate at the speed of light. Hence, time
  evolution equations must be solved. This is done by splitting 4D
  spacetime into 3D spatial slices that are evolved in the time
  direction.  In the simplest way of writing the equations (the
  so-called Arnowitt-Deser-Misner [ADM] formulation), they form a
  system of 12 partial differential evolution equations, 4 gauge
  variables that must be specified (and evolved in time or
  recalculated on each slice), and 4 elliptic \emph{constraint}
  equations without time derivatives. The ADM formulation has poor
  numerical stability properties. These lead to violations of the
  constraint equations and numerical instabilities that make long-term
  evolution impossible.

  \hspace{1em} It took until the 2000s for numerical relativity to
  find formulations of Einstein's equations and gauge choices that
  together lead to stable long-term evolutions. In some cases,
  well-posedness and strong or symmetric hyperbolicity can be
  proven. The equations are typically evolved time-explicitly with
  straightforward high-order (fourth and higher) finite difference
  schemes or with multi-domain pseudospectral methods.

  \hspace{1em} Since numerical relativity only recently became
  applicable to astrophysical simulations, very few core-collapse
  supernova codes are fully general relativistic at this point
  \cite{kuroda:16,ott:13a}. The fully general-relativistic approach is
  much more memory and FLOP intensive than
  solving the Newtonian Poisson equation. Its advantage in large-scale
  computations, however, is the hyperbolic nature of the equations,
  which does not require global matrix inversions or summations and
  thus is advantageous for the parallel scaling of the algorithm.

  \vspace*{0.5em}
  
\item \textbf{Neutrino Transport and Neutrino-Matter Interactions.}
  Neutrinos move at the speed of light (the very small neutrino masses
  are neglected) and can travel macroscopic distances between
  interactions.  Therefore, they must be treated as non-equilibrium
  radiation. Radiation transport is closely related to kinetic
  theory's Boltzmann equation. It describes the phase-space evolution
  of the neutrino distribution function or, in radiation transport
  terminology, their specific intensity. This is a 6+1 D problem: 3
  spatial dimensions, neutrino energy, and two momentum space
  propagation angles in addition to time. The angles describe the
  directions from which neutrinos are coming and where they are going
  at a given spatial coordinate. In addition, the transport equation
  must be solved separately for multiple neutrino species: electron
  neutrinos, electron antineutrinos, and heavy-lepton ($\mu$, $\tau$)
  neutrinos and antineutrinos.
  
  \hspace{1em} Figure~\ref{fig:momspace} shows map projections of the
  momentum space angular neutrino distribution at different radii in a
  supernova core. In the dense protoneutron star, neutrinos are
  trapped and in equilibrium with matter. Their radiation field is
  isotropic. They gradually diffuse out and decouple from matter at
  the \emph{neutrinosphere} (the neutrino equivalent of the
  photosphere). This decoupling is gradual and marked by the
  transition of the angular distribution into the forward (radial)
  direction. In the outer decoupling region, neutrino heating is
  expected to occur and the heating rates are sensitive to the angular
  distribution of the radiation field (cf. \cite{ott:08}).
  Eventually, at radii of a few hundred kilometers, the neutrinos have
  fully decoupled and are free streaming. Neutrino interactions with
  matter (and thus the decoupling process) are very sensitive to
  neutrino energy, since weak-interaction cross-sections scale with
  the square of the neutrino energy. This is why neutrino transport
  needs to be \emph{multi-group}, with typically a minimum of $10-20$
  energy groups covering supernova neutrino energies of $1 -
  \mathcal{O}(100)\,\mathrm{MeV}$. Typical mean energies of electron
  neutrinos are around $10-30\,\mathrm{MeV}$.  Energy exchanges
  between matter and radiation occur via the collision terms in the
  Boltzmann equation. These are stiff sources/sinks that must be
  handled time-implicitly with (local) backward-Euler methods.  The
  neutrino energy bins are coupled through (1) frame-dependent energy
  shifts since the material neutrinos interact with is moving, (2)
  gravitational redshift, and (3) energy transfer in scatterings off
  of electrons and nucleons. Neutrino-matter interaction rates are
  usually precomputed and stored in dense multi-D tables within which
  simulations interpolate.

  \hspace{1em} Full 6+1~D general-relativistic Boltzmann
  neutrino-radiation hydrodynamics is exceedingly challenging and has
  so far not been possible to included in core-collapse supernova
  simulations. 3+1~D (1D in space, 2D in momentum space)
  (e.g., \cite{liebendoerfer:05}), 5+1~D (2D in space, 3D in momentum space)
  simulations \cite{ott:08} and static 6D simulations
  \cite{sumiyoshi:15} have been carried out.

  \hspace{1em} Most (spatially) multi-D simulations treat neutrino
  transport in some dimensionally-reduced approximation. The most
  common is an expansion of the radiation field into angular
  moments. The $n$-th moment of this expansion requires information
  about the $(n+1)$-th moment (and in some cases also about the
  $(n+2)$-th moment). This necessitates a closure relation for the
  moment at which the expansion is truncated. Multi-group flux-limited
  diffusion evolves the $0$-th moment (the radiation energy
  density). The flux limiter is the closure that interpolates between
  diffusion and free streaming. The disadvantages of this method are
  its very diffusive nature that washes out spatial variations of the
  radiation field, its sensitivity to the choice of flux limiter, and
  the need for time-implicit integration (involving global matrix
  inversion) due to the stability properties of the parabolic
  diffusion equation. Two-moment transport is the next better
  approximation. It solves equations for the radiation energy density
  and momentum (i.e. the radiative flux) and requires a closure that
  describes the radiation pressure tensor (also known as the Eddington
  tensor). This closure can be analytic and based on the local values
  of energy density and flux (the M1 approximation). Alternatively,
  some codes compute a global closure based on the solution of a
  simplified, time-independent Boltzmann equation. The major advantage
  of the two-moment approximation is that its advection terms are
  hyperbolic and can be handled with standard time-explicit
  finite-volume methods of computational hydrodynamics and only the
  local collision terms need time-implicit updates.

  \hspace{1em} There are now implementations of multi-group two-moment
  neutrino radiation-hydrodynamics in multiple 2D/3D core-collapse
  supernova simulation codes (e.g., \cite{oconnor:15b,kuroda:16,roberts:16b}).
  This method may be sufficiently close to the full Boltzmann solution
  (in particular if a global closure is used) and appears to be the
  way toward massively-parallel long-term 3D core-collapse supernova
  simulations.

  \vspace*{0.5em}
  
  \item \textbf{Neutrino Oscillations.} Neutrinos have mass and can
    oscillate between flavors. The oscillations occur in vacuum, but
    can also be mediated by neutrino-electron scattering (the
    Mikheyev-Smirnov-Wolfenstein [MSW] effect) and neutrino-neutrino
    scattering. Neutrino oscillations depend on neutrino mixing
    parameters and on the neutrino mass eigenstates (the magnitudes of
    the mass differences are known, but not their signs). Observation
    of neutrinos from the next galactic core-collapse supernova could
    help constrain the neutrino mass hierarchy (see the recent review
    by \cite{mirizzi:16}).

    \hspace{1em} MSW oscillations occur in the stellar envelope. They
    are important for the neutrino signal observed in detectors on
    Earth, but they cannot influence the explosion itself. The
    self-induced (via neutrino-neutrino scattering) oscillations,
    however, occur at the extreme neutrino densities near the
    core. They offer a rich phenomenology that includes collective
    oscillation behavior of neutrinos (see the review in
    \cite{mirizzi:16}).  The jury is still out on their potential
    influence on the explosion mechanism.

    \hspace{1em} Collective neutrino oscillation calculations
    (essentially solving coupled Schr\"odinger-like equations) are
    computationally intensive \cite{mirizzi:16}. They are currently
    performed independently of core-collapse supernova simulations and
    do not take into account feedback on the stellar plasma. Fully
    understanding collective oscillations and their impact on the
    supernova mechanism will quite likely require that neutrino
    oscillations, transport, and neutrino-matter interactions are solved
    for together in a quantum-kinetic approach \cite{vlasenko:14a}.

    \vspace*{0.5em}
  
\item \textbf{Equation of State and Nuclear Reactions.}  The EOS is
  essential for the (M)HD part of the problem and for updating the
  matter thermodynamics after neutrino-matter interactions. Baryons
  (proton, neutrons, alpha particles, heavy nuclei), electrons,
  positrons, and photons contribute to the EOS. Neutrino momentum
  transfer contributes an effective pressure that is taken into
  account separately since neutrinos are not everywhere in local
  thermodynamic equilibrium with the stellar plasma. In different
  parts of the star, different EOS physics applies.

  \hspace{1em} At low densities and temperatures below
  $\sim$$0.5\,\mathrm{MeV}$ ($\sim$$5\times 10^9\,\mathrm{K}$),
  nuclear reactions are too slow to reach nuclear statistical
  equilibrium. In this regime, the mass fractions of the various heavy
  nuclei (isotopes, in the following) must be tracked explicitly. As
  the core collapses, the gas heats up and nuclear burning must be
  tracked with a nuclear reaction network, a stiff system of
  ODEs. Solving the reaction network requires the inversion of sparse
  matrices at each grid point.  Depending on the number of isotopes
  tracked (ranging, typically from $\mathcal{O}(10)$ to
  $\mathcal{O}(100)$), nuclear burning can be a significant
  contributor to the overall computational cost of a simulation.  The
  EOS in the burning regime is simple, since all isotopes can
  essentially be treated as non-interacting ideal Boltzmann gases.
  Often, corrections for Coulomb interactions are included.  Photons
  and electrons/positrons can be treated everywhere as ideal Bose and
  Fermi gases, respectively. Since electrons will be partially or
  completely degenerate, computing the electron/positron EOS involves
  the FLOP-intensive solution of Fermi integrals.  Because of this,
  their EOS is often included in tabulated form.

  \hspace{1em} At temperatures above $\sim$$0.5\,\mathrm{MeV}$,
  nuclear statistical equilibrium holds. This greatly simplifies
  things, since now the electron fraction $Y_e$ (number of electrons
  per baryon; because of macroscopic charge neutrality, $Y_e$ is equal
  to $Y_p$, the number fraction of protons) is the only compositional
  variable.  The mass fractions of all other baryonic species can be
  obtained by solving Saha-like equations for compositional
  equilibrium.  At densities below $\sim$$10^{10} -
  10^{11}\,\mathrm{g\,cm}^{-3}$ the baryons can still be treated as
  ideal Boltzmann gases (but including Coulomb corrections).

  \hspace{1em} The nuclear force becomes relevant at densities near
  and above $10^{10} - 10^{11}\,\mathrm{g\,cm}^{-3}$. It is an
  effective quantum many-body interaction of the strong force and its
  detailed properties are presently not known.  Under supernova
  conditions, matter will be in NSE in the nuclear regime and the EOS
  is a function of density, temperature, and $Y_e$. Starting from a
  nuclear force model, an EOS can be obtained in multiple ways (see
  the \cite{steiner:13b} for an overview discussion), including direct
  Hartree-Fock many-body calculations, mean field models, or
  phenomenological models (e.g., the liquid-drop model). Typically,
  the minimum of the Helmholtz free energy is sought and all
  thermodynamic variables are obtained from derivatives of the free
  energy. In most cases, EOS calculations are too time consuming to be
  performed during a simulation. As in the case of the
  electron/positron EOS, large ($\gtrsim$200\,MB; must be stored by
  each \texttt{MPI} process), densely spaced nuclear EOS tables are
  precomputed and simulations efficiently interpolate in $(\log \rho,
  \log T, Y_e)$ to obtain thermodynamic and compositional information.

\end{itemize}

\begin{figure}[t]
  \centering
\includegraphics[width=\columnwidth]{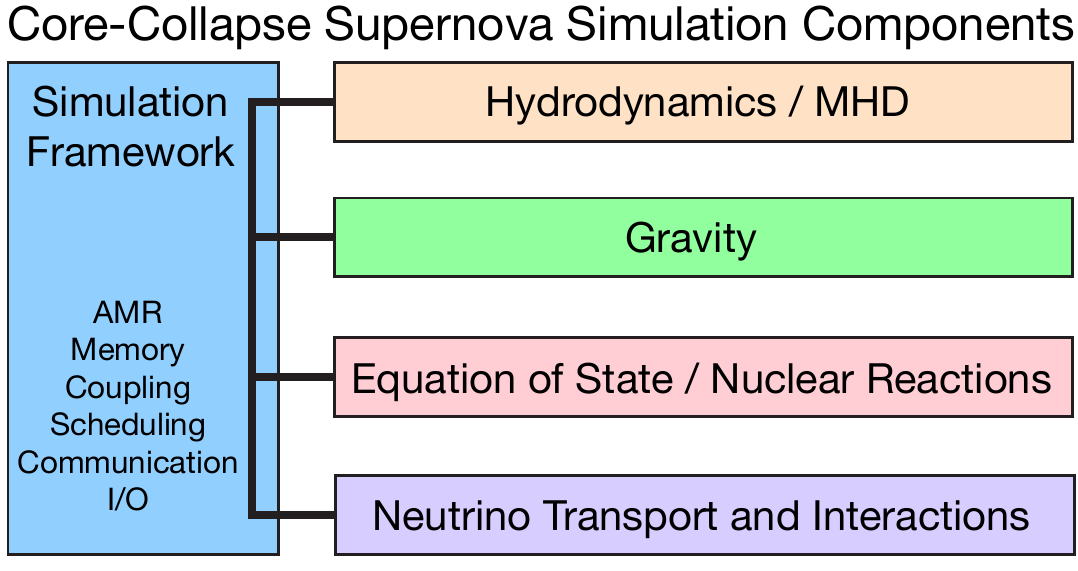}
\caption{Multi-physics modules of core-collapse supernova simulation
  codes. The simulation framework provides parallelization, I/O,
  execution scheduling, AMR, and memory management.}
\label{fig:components}
\end{figure}

\subsection{Effects of Multidimensionality}

Stars are, at zeroth order, gas spheres. It is thus natural to start
with assuming spherical symmetry in simulations -- in particular given
the very limited compute power available to the pioneers of supernova
simulations. After decades of work, it appears now clear that detailed
spherically symmetric simulations robustly
fail at producing explosions for stars that are observed to explode in
nature. Spherical symmetry itself may be the culprit, since symmetry
is clearly broken in core-collapse supernovae:

\hspace{0.5em} (\emph{i}) Observations show that neutron stars receive
``birth kicks'' giving them typical velocities of
$\mathcal{O}(100)\,\mathrm{km\,s}^{-1}$ with respect to the center of mass
of their progenitors. The most likely and straightforward explanation
for these kicks are highly asymmetric explosions leading to neutron
star recoil owing to momentum conservation.

\hspace{0.5em} (\emph{ii}) Deep observations of supernova remnants 
show that the innermost supernova ejecta exhibit low-mode asphericity
similar to the geometry of the shock front shown in
Figure~\ref{fig:full3d}.

\hspace{0.5em} (\emph{iii}) Analytic considerations and also 1D
core-collapse simulations show that the protoneutron star and the
region behind the stalled shock where neutrino heating takes place are
both unstable to buoyant convection, which always leads to the
breaking of spherical symmetry.

\hspace{0.5em} (\emph{iv}) Rotation and magnetic fields naturally
break spherical symmetry. Observations of young pulsars show that some
neutron stars must be born with rotation periods of order
$10\,\mathrm{milliseconds}$. Magnetars may be born with even shorter
spin periods if their magnetic field is derived from rapid
differential rotation.

\hspace{0.5em} (\emph{v}) Multi-D simulations of the violent nuclear
burning in the shells overlying the iron core show that large-scale
deviations from sphericity develop that couple into the precollapse
iron core via the excitation of non-radial pulsations
\cite{couch:15b}. These create perturbations from which convection
will grow after core bounce.

Given the above, multi-D simulations are essential for studying the
dynamics of the supernova engine.

The rapid increase of compute power since the early 1990s has
facilitated increasingly detailed 2D radiation-hydrodynamics
simulations over the past two and a half decades. 3D simulations with
simplified neutrino treatments have been carried out since the early
2000s. The first 3D neutrino radiation-hydrodynamics simulations have
become possible only in the past few years, thanks to the compute
power of large petascale systems like US NSF/NCSA Blue Waters, US
DOE/ORNL Titan or the Japanese K computer.

\section{Core-Collapse Supernova Simulation Codes}

Many 1D codes exist, some are no longer in use, and one is open source
and free to download (\url{http://GR1Dcode.org}). There are $\sim$$10$
(depending on how one counts) multi-D core-collapse supernova
simulation codes in the community.  Many, in particular the 3D codes,
follow the design encapsulated by Figure~\ref{fig:components}. They
employ a simulation framework (e.g., \texttt{FLASH},
\url{http://flash.uchicago.edu/site/flashcode/} or \texttt{Cactus}
\url{http://cactuscode.org}) that handles domain decomposition, 
message passing, memory management, AMR, coupling of different physics
components, execution scheduling, and I/O.

Given the tremendous memory requirement and FLOP-consumption of the
core-collapse supernova problem, these codes are massively parallel
and employ both node-local \texttt{OpenMP} and inter-node \texttt{MPI}
parallelization. All current codes follow a data-parallel paradigm
with monolithic sequential scheduling. This limits scaling, can create
load imbalances with AMR, and makes the use of GPU/MIC accelerators
challenging, since communication latencies between accelerator and CPU
block execution in the current paradigm.

The Caltech \texttt{Zelmani} \cite{ott:13a} core collapse
simulation package is an example of a 3D core-collapse supernova
code. It is based on the open-source \texttt{Cactus} framework, uses
3D AMR Cartesian and multi-block grids, and employs many components
provided by the open-source \texttt{Einstein Toolkit}
(\url{http://einsteintoolkit.org}). \texttt{Zelmani} has fully
general-relativistic gravity and implements general-relativistic
MHD. Neutrinos are included either via a rather crude energy-averaged
leakage scheme that approximates the overall energetics of neutrino
emission and absorption or via a general-relativistic two-moment M1
radiation-transport solver that has recently been deployed on first
simulations \cite{roberts:16b}.

In full radiation-hydrodynamics simulations of the core-collapse
supernova problem with 8 levels of AMR, \texttt{Zelmani}
exhibits good strong scaling with hybrid-\texttt{OpenMP/MPI} to
$16,000$ cores on NSF/NCSA Blue Waters. At larger core counts, load
imbalances due to AMR prolongation and synchronization operations begin
to dominate the execution time. 

\begin{figure}[t]
  \centering
\includegraphics[width=1.0\columnwidth]{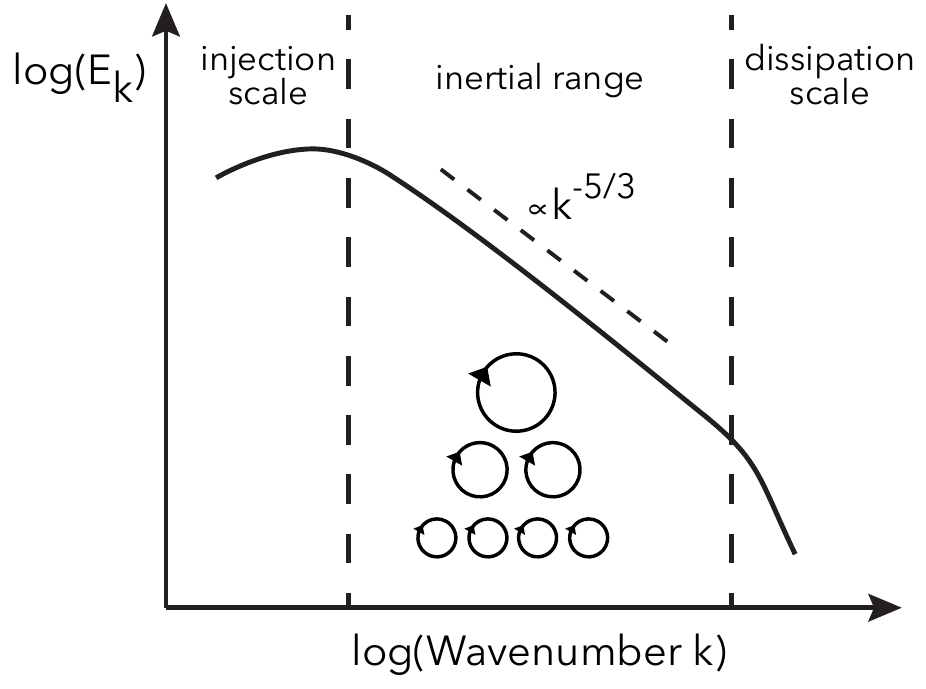}
\caption{Schematic view of turbulence: kinetic energy is injected into
  the flow at large scales and cascades through the inertial range via
  non-linear interactions of turbulent eddies to small scales (high
  wavenumbers in the spectral domain) where it dissipates into
  heat. The scaling of the turbulent kinetic energy with wavenumber in
  the inertial range is $\propto k^{-5/3}$ for Kolmogorov
  turbulence. This scaling is also found in very high-resolution simulations
  of neutrino-driven convection \cite{radice:16a}.}
\label{fig:turbschematic}
\end{figure}

\begin{figure*}[t]
  \centering
\includegraphics[width=1.75\columnwidth]{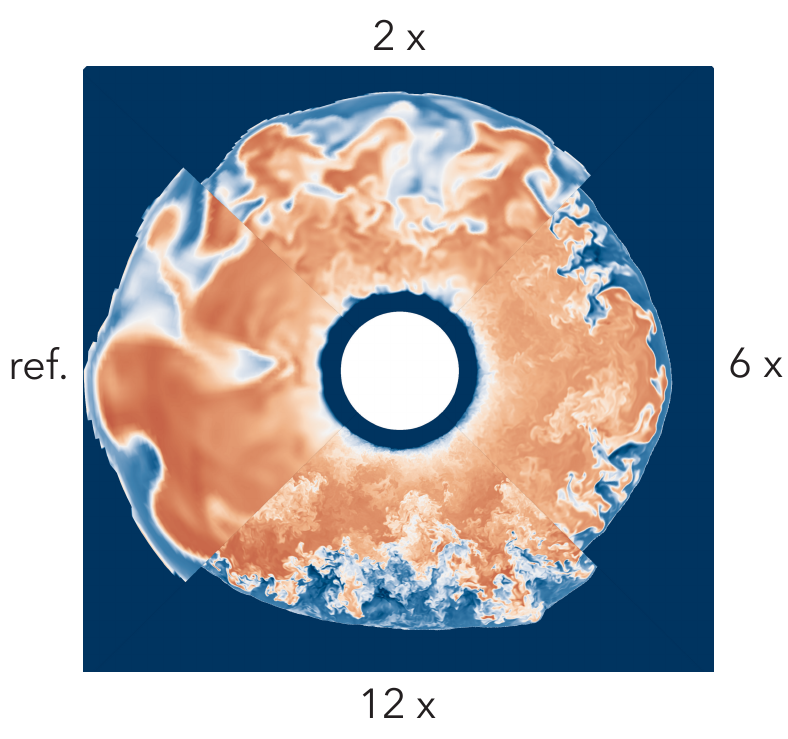}
\caption{Slices from four semi-global 3D simulations of
  neutrino-driven convection with parameterized neutrino cooling and
  heating, carried out in a 45$^\circ$ wedge. The
  colormap is the specific entropy; blue colors mark low entropy
  region, red colors correspond to high entropy. Only the resolution
  is varied. The wedge marked ``ref.'' is the reference resolution
  ($\Delta r = 3.8\,\mathrm{km}$, $\Delta \theta = \Delta \varphi =
  1.8^\circ$) that corresponds to the resolution of present global
  3D detailed radiation-hydrodynamics core-collapse supernova
  simulations. Note how low resolution favors large flow features and
  how the turbulence breaks down to progressively smaller features
  with increasing resolution. This figure uses simulation results of
  \cite{radice:16a} that includes simulations up to 12 times the reference
  resolution that were run on $65,536$ cores of NSF/NCSA Blue Waters. Rendering
  by David Radice (Caltech).}
\label{fig:turb}
\end{figure*}

\section{Multi-D Dynamics and Turbulence}

Even before the first detailed 2D simulations of neutrino-driven
core-collapse supernovae became possible in the mid 1990s, it was
clear that buoyant convection in the protoneutron star and in the
neutrino-heated region just behind the stalled shock breaks spherical
symmetry. Neutrino-driven convection is due to a negative radial
gradient in the specific entropy, making the plasma at smaller radii
``lighter'' than overlying plasma. This is a simple consequence of
neutrino heating being strongest at the base of the heating
region. Rayleigh-Taylor-like plumes develop from small perturbations
and grow to non-linear convection. This convection is extremely
turbulent, since the physical viscosity in the heating region is
vanishingly small. Neutrino-driven turbulence is anisotropic on large
scales (due to buoyancy), mildly compressible (the flow reaches Mach
numbers of $\sim$$0.5$), and only quasi-stationary, because eventually
an explosion develops. Nevertheless, it turns out that Kolmogorov's
description for isotropic, stationary, incompressible turbulence works
surprisingly well for neutrino-driven turbulence (see
Figure~\ref{fig:turbschematic} for a schematic description of
Kolmogorov turbulence and \cite{radice:16a}).

There is something special about neutrino-driven convection in
core-collapse supernovae: unlike convection in globally hydrostatic
stars, neutrino-driven convection occurs on top of a downflow of outer
core material that has accreted through the stalled shock and is
headed for the protoneutron star. The consequence of this is that
there is a competition between (\emph{i}) the time it takes for a
small perturbation to grow to macroscopic scale to become buoyant and
(\emph{ii}) the time it takes for it to leave the region that is
convectively unstable (the heating region) as it is dragged with the
background flow toward the protoneutron star. This means that there
are three parameters governing the appearance of neutrino-driven
convection: the strength of neutrino heating, the initial size of
perturbations entering through the shock, and the downflow rate
through the heating region.  Because of this, neutrino-driven
convection is not a given and simulations find that it does not
develop in some stars.

Even in the absence of neutrino-driven convection, there is another
instability that breaks spherical symmetry in the supernova core: the
standing accretion shock instability (SASI, \cite{janka:12a}). SASI
was first discovered in simulations that did not include neutrino
heating. It works via a feedback cycle: small perturbations enter
through the shock, flow down to the protoneutron star and get
reflected as sound waves that in turn perturb the shock. The SASI is a
low-mode instability that is most manifest in an up-down sloshing
($\ell = 1$ in terms of spherical harmonics) along the symmetry axis
in 2D and in a spiral mode ($m = 1$) in 3D. Once it has reached
non-linear amplitudes, the SASI creates secondary shocks (entropy
perturbations) and shear flow from which turbulence develops.  SASI
appears to dominate in situations in which neutrino-driven convection
is weak or absent: in conditions where neutrino heating is weak, the
perturbations entering the shock are small, or the downflow rate
through the heating region is high.

Independent of how spherical symmetry is broken in the heating region,
all simulations agree that 2D/3D is much more favorable for explosion
than 1D. Some 2D and 3D simulations yield explosions for stars where
1D simulations fail (see, e.g., \cite{lentz:15}). Why is that?

There are two reasons. The first reason has been known for long and is
seemingly trivial: the added degrees of freedom, lateral motion in 2D,
and lateral and azimuthal motion in 3D, have the consequence that a
gas element that enters through the shock front spends more time in
the heating region before flowing down to settle onto the protoneutron
star. Since it spends more time in the heating region, it can absorb
more neutrino energy, increasing the overall efficiency of the
neutrino mechanism.

The second reason has to do with turbulence and has become apparent
only in the past few years. Turbulence is often analyzed employing
\emph{Reynolds decomposition}, a method that separates background flow
from turbulent fluctuations. Using this method, one can show that
turbulent fluctuations lead to an effective dynamical ram pressure
(\emph{Reynolds stress}) that contributes to the overall momentum
balance between behind and in front of the stalled shock. The
turbulent pressure is available only in 2D/3D simulations and it has
been demonstrated (see, e.g., \cite{couch:15a}) that because of this
pressure, 2D/3D core-collapse supernovae explode with less thermal
pressure, and, consequently with less neutrino heating.

Now, the Reynolds stress is dominated by turbulent fluctuations at the
largest physical scales: A simulation that has more kinetic energy in
large-scale motions will explode more easily than a simulation that
has less. This realization readily explains recent findings by
multiple simulation groups: 2D simulations appear to explode more
readily than 3D simulations \cite{couch:15a,lentz:15}. This is likely
a consequence of the different behaviors of turbulence in 2D and
3D. In 2D, turbulence transports kinetic energy to large scales (which
is unphysical), artificially increasing the turbulent pressure
contribution. In 3D, turbulence cascades energy to small scales (as it
should and is known experimentally), so a 3D supernova will generally
have less turbulent pressure support than a 2D supernova.

Another recent finding by multiple groups is that simulations with
lower spatial resolution appear to explode more readily than
simulations with higher resolution. There are two possible
explanations for this and it is likely that they play hand-in-hand:
(\emph{1}) Low resolution creates a numerical bottleneck in the
turbulent cascade, artificially trapping turbulent kinetic energy at
large scales where it can contribute most to the explosion. (\emph{2})
Low resolution also increases the size of numerical perturbations that
enter through the shock and from which buoyant eddies form. The larger
these seed perturbations are, the stronger is the turbulent convection
and the larger is the Reynolds stress.
  
The qualitative and quantitative behavior of turbulent flow is very
sensitive to numerical resolution. This can be appreciated by looking
at Figure~\ref{fig:turb}, which shows the same 3D simulation of
neutrino-driven convection at 4 different resolutions, spanning a
factor of 12 from the reference resolution that is presently used in
many 3D simulations and which underresolves the turbulent flow. As
resolution is increased, turbulent flow breaks down to progressively
smaller features. What also occurs, but cannot be appreciated from a
still figure, is that the intermittency of the flow increases as the
turbulence is better resolved. This means that flow features are not
persistent, but quickly appear and disappear through non-linear
interactions of turbulent eddies. In this way, the turbulent cascade
can be temporarily reversed (this is called \emph{backscatter} in
turbulence jargon), creating large-scale intermittent flow features
similar to what is seen at low resolution. The role of intermittency
in neutrino-driven turbulence and its effect on the explosion
mechanism remain to be studied.

A key challenge for 3D core-collapse supernova simulations is to
provide sufficient resolution so that kinetic energy cascades away
from the largest scales at the right rate. Resolution studies suggests
that this may require between twice to ten times the resolution of
current 3D simulations \cite{radice:16a}.  A ten-fold increase in
resolution in 3D corresponds to a $10,000$ times increase in the
computational cost. An alternative may be to devise an efficient
\emph{sub-grid} model that, if included, provides for the correct rate
of energy transfer to small scales.  Work in that direction is still
in its infancy in the core-collapse supernova context.


\section{Making Magnetars:\\ Resolving the Magnetorotational Instability}

\begin{figure}[t]
  \centering
\includegraphics[width=0.8\columnwidth]{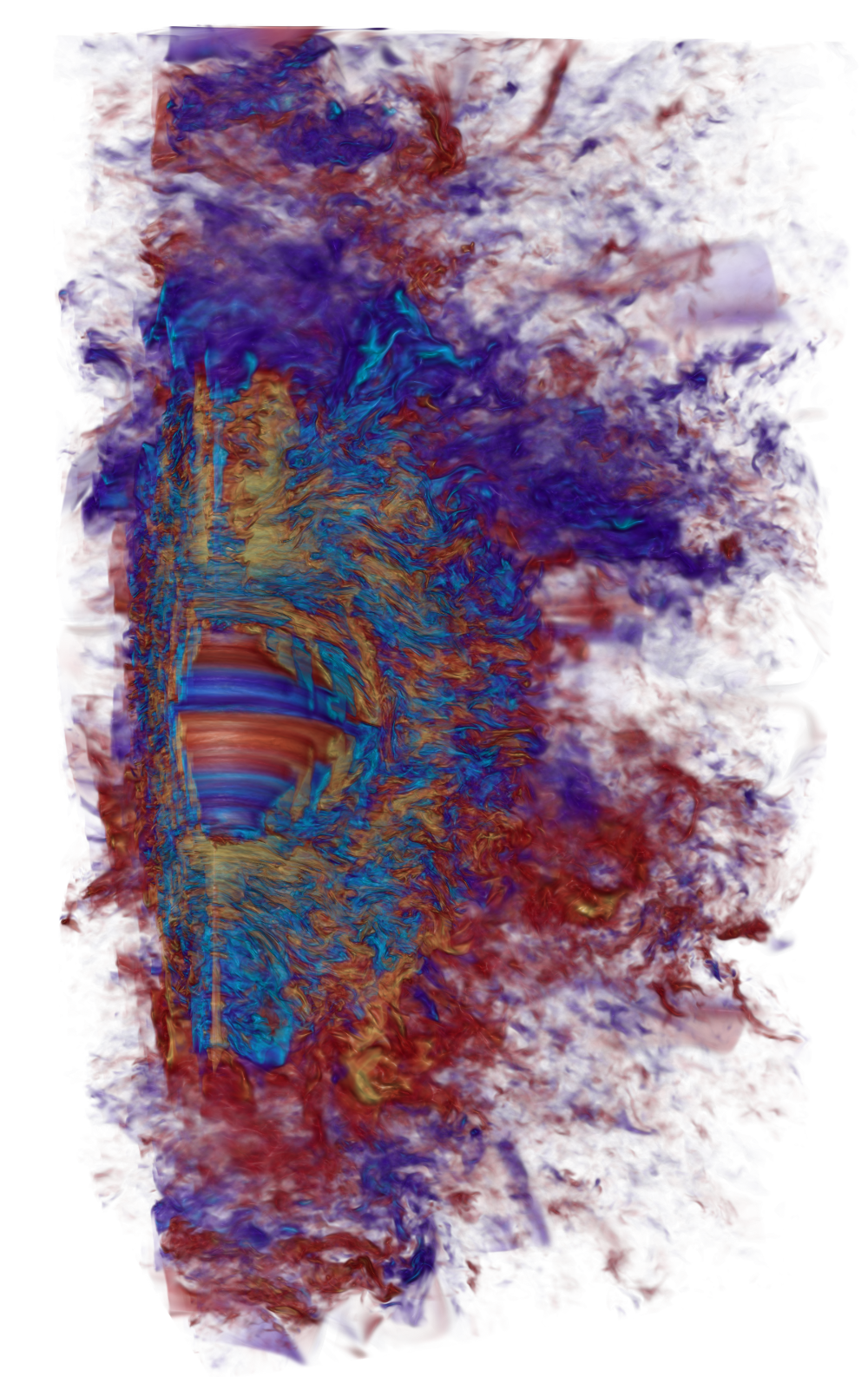}
\caption{Visualization of the toroidal magnetic field built up by an
  inverse cascade (large-scale dynamo) from small-scale
  magnetoturbulence in a magnetorotational core-collapse
  supernova. Shown is a $140 \times 70\,\mathrm{km}$ 3D \emph{octant}
  region with periodic boundaries on the $x-z$ and $y-z$ faces.  Regions
  of strongest positive and negative magnetic field are marked by
  light blue and yellowish colors. Dark blue and dark red colors mark
  regions of weaker negative and positive magnetic field.  Based on
  the NSF/NCSA Blue Waters simulations of M\"osta~\emph{et al.}~2015
  \cite{moesta:15} and rendered by Robert R. Sisneros (NCSA) and
  Philipp M\"osta (UC Berkeley).}
\label{fig:mhd}
\end{figure}

The magnetorotational mechanism relies on the presence of an
ultra-strong ($\sim$$10^{15}-10^{16}\,\mathrm{G}$) global, primarily
toroidal, magnetic field around the protoneutron star. Such a strongly
magnetized protoneutron star is called a \emph{protomagnetar}.

It has been theorized that the magnetorotational instability (MRI,
\cite{balbus:91}) could generate a strong local magnetic field that
could be transformed into a global field by a dynamo process. While
appealing, it was not at all clear that this is what happens.  The
physics is fundamentally global and 3D and global 3D MHD simulations
with sufficient resolution to capture MRI-driven field growth were
impossible to perform for core-collapse supernovae.

This changed with the advent of Blue Waters-class petascale
supercomputers and is a testament to how increased compute power and
capability systems like Blue Waters facilitate scientific
discovery. In M\"osta~\emph{et al.} 2015 \cite{moesta:15}, our group
at Caltech carried out full-physics 3D global general-relativistic MHD
simulations of ten milliseconds of a rapidly spinning protoneutron
star's life, starting shortly after core bounce. We cut out a central
octant (with appropriate boundary conditions) from another,
lower-resolution 3D AMR simulation, and covered a 3D region of $140
\times 70 \times 70\,\mathrm{km}$ with uniform resolution.  We
performed four simulations to study the MHD dynamics at resolutions of
$500\,\mathrm{m}$ ($\sim$2 points per MRI wavelength),
$200\,\mathrm{m}$, $100\,\mathrm{m}$, and $50\,\mathrm{m}$ ($\sim$20
points per MRI wavelength). Since we employed uniform resolution and
no AMR, the simulations showed excellent strong scaling. The
$50\,\mathrm{m}$ simulation was run on $130,000$ Blue Waters cores. It
consumed roughly $3\,\mathrm{million}$ Blue Waters node hours
($\sim$48\,million CPU hours).

Our simulations with $100\,\mathrm{m}$ and $50\,\mathrm{m}$ resolution
resolve the MRI and show exponential growth of the magnetic field.
This growth saturates at small scales within a few milliseconds and is
consistent with what one anticipates on the basis of analytical
estimates. The MRI drives MHD turbulence that is most prominent in the
layer of greatest rotational shear, just outside of the protoneutron
star core at radii of $20-30\,\mathrm{km}$. What we did not anticipate
is that in the highest-resolution simulation (which resolves the
turbulence best), an inverse turbulent cascade develops that
transports magnetic field energy toward large scales. It acts as a
large-scale dynamo that builds up global, primarily toroidal field,
just in the way needed to power a magnetorotational
explosion. Figure~\ref{fig:mhd} shows the final toroidal magnetic
field component in our $50\,\mathrm{m}$ simulation after
$10\,\mathrm{ms}$ of evolution time. Regions of strongest positive and
negative magnetic field are marked by yellowish and light blue colors,
respectively, and are just outside the protoneutron star core. At the
time shown, the magnetic field on large scales has not yet reached its
saturated state. We expect this to occur after
$\sim$$50\,\mathrm{ms}$, which could not be simulated.

The results of M\"osta~\emph{et al.} suggest that the conditions
necessary for the magnetorotational mechanism are a generic outcome of
the collapse of rapidly rotating cores. The MRI is a weak field
instability and will grow to the needed saturation field strengths
from any small seed magnetic field. The next step is to find a
way to simulate for longer physical time and with a larger physical
domain. This will be necessary in order to determine the long-term
dynamical impact of the generated large-scale magnetic field. Such
simulations will require algorithmic changes to improve parallel
scaling, facilitate the efficient use of accelerators, and may require
even larger and faster machines than Blue Waters.

\section{Concluding Remarks}

Core-collapse supernova theorists have always been among the top group
of users of supercomputers. The CDCs and IBMs of the 1960s and 1970s,
the vector Crays of the 1970s to 1990s, the large parallel scalar
architectures of the 2000s, and the current massively parallel SIMD
machines all paved the path of progress for core-collapse supernova
simulations.

Today's 3D simulations are rapidly improving in their included
macroscopic and microscopic physics. They are beginning to answer
decades-old questions and are allowing us to formulate new
questions. There is still much need for improvement, which will come
at no small price in the post-Moore's-law era of heterogeneous
supercomputers.

One important issue that the community must address is the
reproducibility of simulations and the verification of simulation
codes. It still occurs more often than not that different codes
starting from the same initial conditions and implementing nominally
the same physics arrive at quantitatively and qualitatively different
outcomes. In the mid-2000s an extensive comparison of 1D supernova
codes took place that provided results that are still being used as
benchmarks today \cite{liebendoerfer:05}. Efforts are now underway
that will lead to the definition of multi-D benchmarks. In addition to
code comparisons, the increasing availability of open-source
simulation codes and routines for generating input physics (e.g.,
neutrino interactions) is furthering reproducibility. Importantly,
these open-source codes now allow new researchers to enter the field
without the need of spending many years on developing basic simulation
technology that already exists.

Core collapse is, in essence, an initial value problem. Current
simulations, even those in 3D, start from spherically symmetric
precollapse conditions from 1D stellar evolution codes. However, stars
rotate and convection in the layers surrounding the inert iron care is
violently aspherical. These asphericities have an impact on the
explosion mechanism. In order for 3D core-collapse supernova
simulations to provide robust and reliable results, the initial
conditions must be reliable and robust, and will likely require
simulating the final phases of stellar evolution in 3D
\cite{couch:15b}, which is another multi-D, multi-scale, multi-physics
problem.

Neutrino quantum-kinetics for including neutrino oscillations directly
into simulations will be an important, but exceedingly algorithmically
and computationally challenging addition to the simulation
physics. Formalisms for doing so are under development and first
implementations (in spatially 1D) simulations may be available in a few
years.

A single current top-of-the-line 3D neutrino radiation-hydrodynamics
simulation can be carried out to $\sim$$0.5-1\,\mathrm{second}$ after
core bounce at a cost of several tens of millions of CPU hours; and it
still underresolves the neutrino-driven turbulence. What is needed
now, are many such simulations for studying sensitivity to initial
conditions such as rotation and progenitor structure and input
physics. These simulations should be at higher resolution and carried
out for longer so that the longer-term development of the explosion
(or collapse to a black hole) and, for example, neutron star birth
kicks can be reliably simulated.

Many longer simulations at higher resolution will require much more
compute power than is currently available. The good news is that the
next generation of petascale systems and, certainly, exascale machines
in the next decade will provide the necessary FLOPS. The bad news: the
radical and disruptive architectural changes necessary on the route to
exascale will require equally disruptive changes in supernova
simulation codes. Already at petascale, the traditional data-parallel,
linear/sequential execution model of all present supernova codes is
the key limiting factor of code performance and
scaling. A central issue is the need to communicate
  many boundary points between subdomains for commonly employed
  high-order finite difference and finite volume schemes. With
  increasing parallel process count, communication eventually
  dominates over computation in current supernova simulations.

Since latencies cannot be hidden, efficiently offloading data and
tasks to accelerators in heterogeneous systems is difficult for
current supernova codes. The upcoming generation of petascale machines
such as DOE's Summit and Sierra, fully embraces heterogeneity. For
exascale machines, power consumption will be the driver of computing
architecture. Current Blue Waters already draws
$\sim$$10\,\mathrm{MW}$ of power and there is not much upwards
flexibility for future machines. Unless there are unforeseen
breakthroughs in semiconductor technology that provide increased
single-core performance at orders of magnitude lower power footprint,
exascale machines will likely be all-accelerator with hundreds of
millions of slow, highly energy efficient cores.

Accessing the compute power of upcoming petascale and future exascale
machines requires a radical departure from current code design and
major code development efforts. Several supernova groups are exploring
new algorithms, numerical methods, and parallelization
paradigms. Discontinuous Galerkin (DG) finite elements (e.g.,
\cite{hesthaven:07}) have emerged as a promising discretization
approach that guarantees high numerical order while minimizing the
amount of subdomain boundary information that needs to be communicated
between processes. In addition, switching to a new, more flexible
parallelization will likely be necessary to prepare supernova codes
(and other computational astrophysics codes solving similar equations)
for exascale machines. A prime contender being considered by supernova
groups is task-based parallelism, which allows for fine-grained
dynamical load balancing and asynchronous execution and
communication. Frameworks that can become task-based backbones of
future supernova codes already exist, e.g., Charm++
(\url{http://charm.cs.illinois.edu/research/charm}), Legion
(\url{http://legion.stanford.edu/overview/}), and Uintah
(\url{http://uintah.utah.edu/}).

\section*{Acknowledgments}

I acknowledge helpful conversations with and help from Adam Burrows,
Sean Couch, Steve Drasco, Roland Haas, Kenta Kiuchi, Philipp M\"osta,
David Radice, Luke Roberts, Erik Schnetter, Ed Seidel, and Masaru
Shibata. I thank the Yukawa Institute for Theoretical Physics at Kyoto
University for hospitality while writing this article. This work
is supported by NSF under award nos. CAREER PHY-1151197 and TCAN
AST-1333520, and by the Sherman Fairchild Foundation. Computations
were performed on NSF XSEDE under allocation TG-PHY100033 and on
NSF/NCSA BlueWaters under NSF PRAC award no.\ ACI-1440083. Movies of
simulation results can be found on
\url{http://www.youtube.com/SXSCollaboration}.

\ifCLASSOPTIONcaptionsoff
  \newpage
\fi


\end{document}